# Generation and Analysis of Hidden Optical-Polarization States

Ravi S. Singh[1], Gyaneshwar K. Gupta[2]

Department of Physics, D. D. U. Gorakhpur University, Gorakhpur-273009 (U.P.) – INDIA

Email: - [1] yesora27@gmail.com; [2] gyankg@gmail.com

**Abstract:-** A hybrid Optical-Device (Phase-Conjugating Mirror Michelson Interferometer), made up of Phase-conjugate mirror along with ordinary mirror and Polarization Beam-splitter, is formally designed and investigated for the generation of an unusual Optical-Polarization States. This weird yet 'essentially single-mode' optical-polarization state has non-random 'ratio of amplitudes' and non-random 'sum of phases' in orthogonal bases-modes contrary to usual 'truly single-mode' optical-polarization states of which non-random 'ratio of amplitudes' and non-random 'difference of phases' serves as optical-polarization characteristic parameters. Since it is seen not to be characterized by Stokes parameters, one may, therefore, assign the name 'Hidden Optical-Polarization States (HOPS)'. HOPS are characterized by a set of parameters, namely, Hidden Optical-Polarization parameters. Formal experimental schemes are presented to experimentally measure these parameters and, thus, offering characterization of HOPS.

## 1 Introduction

Polarization of light ensures the transversal nature of electromagnetic wave and witnessed seminal contributions from early investigators, notably, C. Huygens, W. Nicole, T. Young, Fresnel, Arago etc. [1]. Classically, optical-polarization is analytically defined as the temporal-evolution of tip of Electric field vector (light vector), which, in general, traverses an ellipse of non-random eccentricity and orientation. On varying 'ratio of amplitudes' and 'difference in phases' in orthogonal bases-modes, serving as characteristic optical-polarization parameters, the ellipse degenerates into linear and circular polarizations. Variant techniques such as Stokes-Parameters, Jones Matrix, and Coherency Matrix [2] have been utilized to quantify Optical-Polarization. Due to viability to the experimental measurements and the application through operatic-modification in Quantum-domain Stokes-Parameters [3] are still affording characterization for Optical-Polarization.

Until 1970's confusion regarding stringent definition of unpolarized light exists. In 1971 Prakash and Chandra [4], and independently Agarwal [5], discovered the structure of density operator for Unpolarized light (natural light) providing its rigorous statistical description. Lehner et al. [6] and others [7-8] re-visited Unpolarized light offering some new insights. Mehta and Sharma [9] defines perfect optical-polarization state by demanding SU(2) transformation in bi-modal monochromatic light into 'truly single-mode' linearly polarized mode. This treatment, while rigorously difining perfect optical-polarization as single-mode states, doesn't prescribe a criterion. Prakash and Singh [10] obtained the optical-polarization operator of which action on any quantum state yields index of polarization, i.e., 'ratio of amplitudes' and 'difference in phases' in orthogonal bases-modes, characteristic optical-polarization parameters.

Coherent states, discovered by Schrodinger [11] and applied cogently in Quantum optics by Glauber, Sudarshan [12-14], offers minimum uncertainties in canonically conjugate variables of optical field. Perelomov [15] introduced and investigated salient features of General Coherent States pertained to representations of an arbitrary Lie group. Since then numerous electromagnetic field-states such as Even and odd coherent states, squeezed states, binomial states, entangled states in variant degrees of freedom, have been proposed and prepared in novel experiments [16-21]. The concept of Squeezing and Entanglement has been borrowed to explicate applications in optical-polarization of electromagnetic radiation. Being members of SU(2) algebra quantum stokes operators are non-commutative observables and, therefore, simultaneous precise measurements are precluded due to Heisenberg Uncertainty principle. Suppression of variance (noise) in one or more stokes operators, below those acquired at vacuum or coherent states, provides signature of Optical-Polarization Squeezing. Grangier et al.[22] generated polarization-squeezed beam to improve the sensitivity of polarization interferometer. Recently,

N. Korolkova et al. [23] formally proposed schemes, which is implemented by Bowen et al. [24] to generate polarization squeezed light and characterize entanglement in Polarization EPR states. Two orthogonal polarization states forming photonic qubit paves the way of novel experiment to test the fundamental issues in Quantum mechanics [25-29] and, on account of being easily manipulated and modulated by linear optical elements, is carrier of quantum information between nodes of quantum network offering linear quantum computing [30-31]. Contrary to aforementioned extensively investigated optical field-states, generalization of Optical-Polarization states is rather less studied and analyzed.

One of the authors and others [32] generalized the concept of optical-polarization by introducing bi-modal monochromatic rectilinearly propagating optical field in which 'ratio of amplitudes' and 'sum in phases', rather than 'ratio of amplitudes' and 'difference in phases' as in usual concept of optical-polarization in transverse bases-modes, are nonrandom characteristic parameters and defined it as 'Hidden Optical-Polarization states (HOPS) of light' [33]. HOPS being mono-modal possesses infinitely many phase-coherent (phase-locked) light whose amplitudes are quite arbitrary but phases preserve two well-defined values. Recently generation of HOPS in Degenerate Parametric Amplification and Squeezing therein is studied if non-linear crystal is pumped by coherent or chaotic light [34-35].

The present work deals with formal proposals for an alternative method of generation of Hidden Optical-Polarization states (HOPS) and measurements for Hidden Optical-Polarization parameters. An experimental set-up is formally designed by imitating Michelson Interferometer in such a way that one of the plane mirrors is replaced by Phase-Conjugating mirror and half-silvered Beam splitter by Polarizing Beam Splitter (PBS). The conceived device may be termed as Phase-Conjugating Michelson Interferometer (PCMI). The paper is arranged in following sections: Section 2 gives a brief account of indices of optical-polarization and hidden optical-polarization which reduces bi-modal description of optical-polarization and that of hidden optical-polarization into mono-modal one. In section 3 design of PCMI and its working principle is outlined for the generation of HOPS. Section 4 deals with inadequacies of Stokes parameter in characterizing HOPS, definitions of Hidden Optical-Polarization parameters and proposals for their measurement. Conclusion is also drawn.

## 2 Indices of Optical-Polarization and Hidden Optical-Polarization

A monochromatic beam of light (optical field) propagating rectilinearly along z-direction, in Classical optics, is governed by Maxwell's Classical Electromagnetic Theory having Vector Potential (analytic signal),

$$\mathcal{A} = [\hat{\mathbf{e}}_x \underline{A}_x + \hat{\mathbf{e}}_y \underline{A}_y] \, e^{-i\psi}, \tag{1}$$

where $\underline{A}_{x,y} = A_{0x,0y} \exp(i\, \varphi_{x,y})$ are classical complex amplitudes in basis-modes ($\hat{\mathbf{e}}_{x,y}$, **k**), $\psi = \omega t - kz$, **k** (= k $\hat{\mathbf{e}}_z$) is propagation vector, and $\hat{\mathbf{e}}_{x,y,z}$ are respective unit vectors along x-, y-, and z- axes. Obviously, vector potential, $\mathcal{A}$, Eq.(1) and, hence, the optical field is completely specified by its real transverse-amplitudes, $A_{0x,0y}$ and phase-parameters, $\varphi_{x,y}$. These four parameters ($A_{0x,0y}$; $\varphi_{x,y}$) have, in general, random spatio-temporal variations providing bimodal unpolarized optical field as it needs two random complex-amplitudes for its complete statistical characterization. In Quantum Optics the optical field, Eq.(1) is quantized utilizing canonical quantization technique [36] which yields positive-frequency part of Vector Potential operator as,

$$\widehat{\mathcal{A}}^{(+)} = \left(\frac{2\pi}{\omega V}\right)^{1/2} [\hat{\mathbf{e}}_x \hat{a}_x(t) + \hat{\mathbf{e}}_y \hat{a}_y(t)] \, e^{ikz}, \tag{2}$$

and the negative-part of the same is obtained by $\widehat{\mathcal{A}}^{(-)} = \widehat{\mathcal{A}}^{(+)\dagger}$. Here $\hat{a}_{x,y}$ are well-known Bosonic-annihilation operators termed as quantized complex amplitudes, $\omega$ is angular frequency of the optical field and V is the quantization volume of the cavity.

Mehta and Sharma [9] provided stringent definition of polarized light in Quantum Optics by transforming rectilinearly propagating bi-modal monochromatic light to a linearly-polarized single-mode on passing through compensator and/or rotator (SU(2)-transformations). Polarized light so defined may be termed as 'truly' single-mode optical field as the signal is absent in orthogonal mode. The usual (ordinary) polarized light is completely determined either by the pair of non-random 'ratio of amplitudes' and non-random 'difference in phases' in orthogonal basis-modes, ($\hat{e}_{x,y}$, $\bar{k}$) or by a non-random complex parameter defined as Index of polarization(IOP) [10].

The concept of Hidden Optical-Polarization States (HOPS) [32] has been introduced in which signal is, in general, present in all modes but only one complex amplitude suffices for its complete statistical description. HOPS, may, therefore, be termed as 'essential' single-mode optical-field state. Notably, HOPS has non-random 'sum of phases' and non-random 'ratio of real amplitudes' contrary to 'truly' single-mode ordinary polarized optical field where non-random 'difference of phases' and non-random 'ratio of real amplitudes' in orthogonally basis-modes ($\hat{e}_{x,y}$, $\bar{k}$). Besides adopting linear-basis of description ($\hat{e}_x$, $\hat{e}_y$) one may work in a general basis ($\hat{\varepsilon}$, $\hat{\varepsilon}_\perp$). $\hat{\varepsilon}$ is complex unit vector, $\hat{\varepsilon} = \varepsilon_x \hat{e}_x + \varepsilon_y \hat{e}_y$, satisfying normalization condition, $\hat{\varepsilon}^* \cdot \hat{\varepsilon} = |\varepsilon_x|^2 + |\varepsilon_y|^2 = 1$. A unit vector orthogonal to $\hat{\varepsilon}$ is given by complex unit vector $\hat{\varepsilon}_\perp$ satisfying $\hat{\varepsilon}_\perp^* \cdot \hat{\varepsilon}_\perp = |\varepsilon_{\perp x}|^2 + |\varepsilon_{\perp y}|^2 = 1$; $\hat{\varepsilon}_\perp \cdot \hat{\varepsilon}^* = \varepsilon_x^* \cdot \varepsilon_{\perp x} + \varepsilon_y^* \cdot \varepsilon_{\perp y} = 0$, providing $\frac{\varepsilon_{\perp y}}{\varepsilon_{\perp x}} = -\frac{\varepsilon_x^*}{\varepsilon_y^*}$, where dot(.) denotes inner product of cartesian vectors and star(*) implies for complex conjugation. The vector potential, $\mathcal{A}$ of a single-mode polarized optical field in the mode ($\hat{\varepsilon}_0$, $\bar{k}$) is described by,

$$\mathcal{A} = \underline{A}\, e^{-i\psi}, \qquad (3)$$

where $\underline{A} = \hat{\varepsilon}_0 \underline{A}$ is the complex amplitude along $\hat{\varepsilon}_0$. Complex amplitudes of optical-field represented by Eq.(3) in the basis ($\hat{\varepsilon}$, $\hat{\varepsilon}_\perp$) are $\underline{A}_{\hat{\varepsilon}} = (\hat{\varepsilon}^* \underline{A}) = \underline{A}(\hat{\varepsilon}^* \cdot \hat{\varepsilon}_0)$; $\underline{A}_{\hat{\varepsilon}_\perp} = (\hat{\varepsilon}_\perp^* \underline{A}) = \underline{A}(\hat{\varepsilon}_\perp^* \cdot \hat{\varepsilon}_0)$ and one may derive IOP in the basis, ($\hat{\varepsilon}$, $\hat{\varepsilon}_\perp$) as

$$p_{(\varepsilon, \varepsilon_\perp)} = \underline{A}_{\hat{\varepsilon}_\perp}/\underline{A}_{\hat{\varepsilon}} = (\hat{\varepsilon}_\perp^* \cdot \hat{\varepsilon}_0)/(\hat{\varepsilon}^* \cdot \hat{\varepsilon}_0), \qquad (4)$$

which is defined to possess non-random values for usual polarized light. Decomposing complex amplitudes, $\underline{A}_{\hat{\varepsilon}}$ ($\underline{A}_{\hat{\varepsilon}_\perp}$) in terms of real amplitudes $A_{0\hat{\varepsilon}}$($A_{0\hat{\varepsilon}_\perp}$) and phase parameters $\varphi_{\hat{\varepsilon}}$($\varphi_{\hat{\varepsilon}_\perp}$) as $\underline{A}_{\hat{\varepsilon}}$ ($\underline{A}_{\hat{\varepsilon}_\perp}$) = $A_{0\hat{\varepsilon}}$($A_{0\hat{\varepsilon}_\perp}$) exp (i$\varphi_{\hat{\varepsilon}}$($\varphi_{\hat{\varepsilon}_\perp}$)), Eq. (4) yields, (i) non-random 'ratio of real amplitudes', $A_{0\hat{\varepsilon}_\perp}/A_{0\hat{\varepsilon}}$ and, (ii) non-random 'difference in phases', $\varphi_{\hat{\varepsilon}_\perp} - \varphi_{\hat{\varepsilon}}$ in basis-modes of description ($\hat{\varepsilon}$, $\hat{\varepsilon}_\perp$). Thus, usual polarized light is completely determined either by non-random 'ratio of amplitudes' and non-random 'difference in phases' in the orthogonal modes or by a non-random complex parameter, $p_{(\hat{\varepsilon},\hat{\varepsilon}_\perp)}$ defining IOP, Eq.(4) cited above. Parametrizing real amplitudes, $A_{0\hat{\varepsilon}}(A_{0\hat{\varepsilon}_\perp})$ and phases $\varphi_{\hat{\varepsilon}}(\varphi_{\hat{\varepsilon}_\perp})$ by introducing real amplitude and angle parameters $A_0$, $\chi$ and $\Delta$ respectively bearing inequalities $0 \leq A_0$, $0 \leq \chi \leq \pi$ and $-\pi < \Delta \leq \pi$ on Poincare sphere we obtain,

$$A_{0\hat{\varepsilon}} = A_0 \cos\frac{\chi}{2}, A_{0\hat{\varepsilon}_\perp} = A_0 \sin\frac{\chi}{2}, \varphi_{\hat{\varepsilon}} = \overline{\varphi} - \Delta/2, \varphi_{\hat{\varepsilon}_\perp} = \overline{\varphi} + \Delta/2, \qquad (5)$$

where $A_0$ and $\overline{\varphi}$ ($0 \leq \overline{\varphi} \leq 2\pi$) are random parameters. One may derive simple relations between old and new parameters as, $A_0 = (A_{0\hat{\varepsilon}}^2 + A_{0\hat{\varepsilon}_\perp}^2)^{1/2}$, $\chi = 2\tan^{-1}(A_{0\hat{\varepsilon}_\perp}/A_{0\hat{\varepsilon}})$, $2\overline{\varphi} = \varphi_{\hat{\varepsilon}_\perp} + \varphi_{\hat{\varepsilon}}$ and $\Delta = \varphi_{\hat{\varepsilon}_\perp} - \varphi_{\hat{\varepsilon}}$. Insertion of Eq.(5) into Eq.(4), one obtains expression, $p_{(\hat{\varepsilon},\hat{\varepsilon}_\perp)} = \underline{A}_{\hat{\varepsilon}_\perp}/\underline{A}_{\hat{\varepsilon}} = \tan\frac{\chi}{2} e^{i\Delta}$ for index of polarization for usual polarized light in basis of description $(\hat{\varepsilon}, \hat{\varepsilon}_\perp)$ whose modulus provide the non-random 'ratio of real amplitudes' and arg $p_{(\hat{\varepsilon},\hat{\varepsilon}_\perp)}$ gives non-random 'difference in phases' for polarized light.

Defining conditions pertaining to HOPS may be casted, in terms of a non-random complex parameter in the basis $(\hat{\varepsilon}, \hat{\varepsilon}_\perp)$, as,

$$p_{h(\hat{\varepsilon},\hat{\varepsilon}_\perp)} = \underline{A}_{\hat{\varepsilon}_\perp}/\underline{A}^*_{\hat{\varepsilon}} = \tan\frac{\chi_h}{2} e^{i\Delta_h}, \qquad (6)$$

where $\chi_h$ and $\Delta_h$ are non-random angle parameters ($0 \leq \chi_h \leq \pi$ and $-\pi < \Delta_h \leq \pi$), defining Index of Hidden Optical-Polarization (IHOP). Parameterzing real amplitudes and phase parameters in a slightly different manner one obtains,

$$A_{0\hat{\varepsilon}} = A_0 \cos \chi_h/2, A_{0\hat{\varepsilon}_\perp} = A_0 \sin \chi_h/2; \varphi_{\hat{\varepsilon}} = \varphi + \Delta_h/2, \varphi_{\hat{\varepsilon}_\perp} = -\varphi + \Delta_h/2 \qquad (7)$$

here $A_0$ and $\varphi$ are random parameters ($0 \leq A_0$, $0 \leq \varphi \leq 2\pi$) contrary to $A_0$ and $\overline{\varphi}$ (see Eq.(5)) satisfying $A_0 = (A^2_{0\hat{\varepsilon}_\perp} + A^2_{0\hat{\varepsilon}})^{1/2}$, $\chi_h = 2\tan^{-1}(A_{0\hat{\varepsilon}_\perp}/A_{0\hat{\varepsilon}})$ and $2\varphi = -(\varphi_{\hat{\varepsilon}_\perp} - \varphi_{\hat{\varepsilon}})$, $\Delta_h = \varphi_{\hat{\varepsilon}_\perp} + \varphi_{\hat{\varepsilon}}$. The distinction between usual Optical-Polarization and Hidden Optical-Polarization is clearly displayed by random parameters $\bar{\varphi}$ (sum in phases) and $\varphi$ (difference in phases) respectively. The vector potential of HOPS can, then, be written in general basis-modes, $(\hat{\boldsymbol{\varepsilon}}, \hat{\boldsymbol{\varepsilon}}_\perp)$ as,

$$\mathcal{A} = [\hat{\boldsymbol{\varepsilon}} \cos \frac{\chi_h}{2} A_0 e^{i\varphi} e^{i\Delta_h/2} + \hat{\boldsymbol{\varepsilon}}_\perp \sin \chi_h/2\, A_0\, e^{-i\varphi} e^{i\Delta_h/2}]\, e^{-i\psi}. \qquad (8)$$

Obviously, Eq.(8) describes mono-modal optical-field in which 'difference of phases', $\varphi$ in orthogonal modes is random parameter and its properties are described by single complex amplitude ($A_0\, e^{i\varphi}$). One may quantized the Hidden Optical-Polarized field, Eq.(8). The positive-frequency part of Vector potential operator for Hidden optically Polarized field will be,

$$\widehat{\mathcal{A}}^{(+)} = \left(\frac{2\pi}{\omega V}\right)^{1/2} [\hat{\boldsymbol{\varepsilon}}\, \hat{a}^\dagger_{\hat{\boldsymbol{\varepsilon}}}(t) + \hat{\boldsymbol{\varepsilon}}_\perp \hat{a}_{\hat{\boldsymbol{\varepsilon}}_\perp}(t)]\, e^{ikz}, \qquad (9)$$

where $\hat{a}_{\hat{\boldsymbol{\varepsilon}}}(\hat{a}_{\hat{\boldsymbol{\varepsilon}}_\perp})$ are usual quantized complex amplitudes (cf. Eq.2) in basis of description $(\hat{\boldsymbol{\varepsilon}}, \hat{\boldsymbol{\varepsilon}}_\perp)$. Using orthonormality conditions of $\hat{\boldsymbol{\varepsilon}}$ and $\hat{\boldsymbol{\varepsilon}}_\perp$ one may derive relations between quantized complex amplitudes $\hat{a}_{x,y}$ and $\hat{a}_{\hat{\boldsymbol{\varepsilon}},\hat{\boldsymbol{\varepsilon}}_\perp}$ as,

$$\hat{a}_{\hat{\boldsymbol{\varepsilon}}} = \hat{\varepsilon}_x^*\, \hat{a}_x + \hat{\varepsilon}_y^*\, \hat{a}_y \text{ and } \hat{a}_{\hat{\boldsymbol{\varepsilon}}_\perp} = -\hat{\varepsilon}_y\, \hat{a}_x + \hat{\varepsilon}_x\, \hat{a}_y \qquad (10)$$

Moreover, the Glauber coherence functions [12] describe correlation properties of optical-field at any spatio-temporal point, providing interference effects, are given by,

$$\Gamma^{(m_x, m_y, n_x, n_y)} = \text{Tr}[\rho(0)\mathcal{A}_x^{(-)m_x}\mathcal{A}_y^{(-)m_y}\mathcal{A}_x^{(+)m_x}\mathcal{A}_y^{(+)m_y}] \qquad (11)$$

where $\rho(0)$ is density operator describing dynamical state of optical field. Setting the condition on quantized complex amplitudes in Eq.(2),

$$\hat{a}_y(t)\rho(0) = p\, \hat{a}_x(t)\rho(0), \qquad (12)$$

where p is IOP for polarized light, one obtains after inserting Eq.(12) into Eq.(11),

$$\Gamma^{(m_x,m_y,n_x,n_y)} = p^{*m_y}p^{m_y}\Gamma^{(m_x+m_y,0,n_x+n_y,0)}, \quad (13)$$

which describes correlation properties of 'truly' single-mode optical-polarization state. Clearly, coherence function, Eq.(13) is determined by p (IOP) and one of quantized complex amplitudes $â_x(t)$. Notably, since, Eq. (13) gives the coherence function for ordinary polarized light, Eq.(12) may be regarded as quantum analogue of classical criterion $\underline{A}_y = p\ \underline{A}_x$ for perfect optical-polarized field. Similarly, having employed the criterion,

$$â_y(t)\rho(0) = p_h e^{-2i\omega t}\rho(0)â^†_x(t), \quad (14)$$

where $p_h$ is index of Hidden Optical-Polarization, and substituting Eq.(14) into Eq.(11), we get coherence functions for 'essential' single-mode Hidden Optical-Polarization state,

$$\Gamma^{(m_x,m_y,n_x,n_y)} = p_h^{*m_y}p_h^{n_y}\Gamma^{(m_x+m_y,0,n_x+n_y,0)} \quad (15)$$

Obviously, Glauber coherence functions, Eq.(15) are governed by $p_h$ (IHOP) and one of the quantized complex amplitudes $â_x(t)$. On a similar ground the Eq.(14) may be regarded as quantum criterion for HOPS, quantum counterpart of the classical criterion $\underline{A}_y = p_h\ \underline{A}_x^*$. Conclusively one may infer that the concepts of indices of optical-polarization and hidden optical-polarization reduces bi-modal description of optical-polarization into mono-modal if indices are provided.

## 3 Phase-Conjugating Mirror Michelson Interferometer (PCMI) and Generation of HOPS

Michelson Interferometer (M I), offering versatile tool to study interference of light beams, had been applied to refute Ether-Drag hypothesis [37] and, thereby, paving the way to put forth postulate for constancy of speed of light in vacuum by Einstien to propound Special Theory of Relativity [38]. A slight modification in design is proposed by replacing one of the two ordinary mirrors by a Phase-Conjugating mirror (Non-linear optical element) [39] and partially-silvered Beam splitter by a Polarizing beam splitter (PBS) (e. g. anisotropic birefringent crystal). So it may be named as Phase-Conjugating Michelson Interferometer (PCMI). Light from a source, S is allowed to incident on Polarizer, P output-light of which is the linearly polarized light, say, in the basis ($\hat{\varepsilon}$, $\hat{\varepsilon}_\perp$). The polarized light passes through PBS (I) which separated spatially the orthogonal polarized components (ray 1↕ ; ray 2 ⊕ ). The ray 1↕ travelling towards ordinary mirror, M, suffers ordinary reflection from it and comes back to PBS (I) and transmits it, to reach at PBS (II). Similarly, the ray 2 ⊕ , while returning back from Phase-Conjugating mirror, bears conjugated-phase with unaltered amplitude (i. e., phase conversion from ωt – kz – φ to ωt + kz + φ + constant), reflects from PBS (I) to superpose with ray 1↕ at PBS (II). The output of PBS (II) is obviously in HOPS, as explained below.

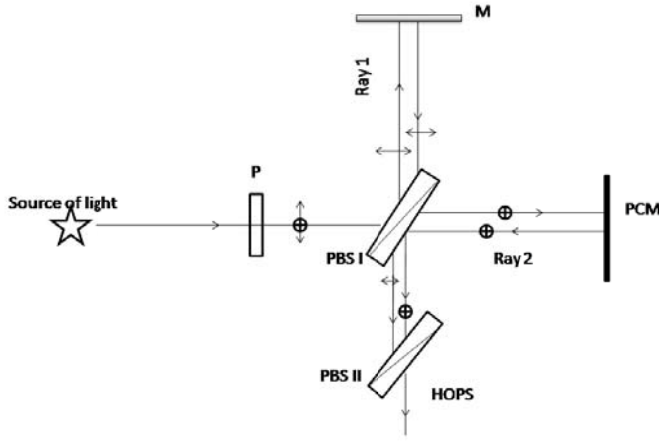

**Fig.1** Formal Experimental Setup for generating HOPS through PCMI; P(polarizer), M (ordinary mirror), PBS(polarizing beam splitter), PCM(phase conjugating mirror), HOPS(hidden optical-polarization states ).

The working principle (Classical theory) of PCMI can be outlined by describing linearly polarized light (output at P) by two orthogonal components ray 1 ↕ and ray 2⊕ with signal-wave in the basis ($\hat{\varepsilon}$, $\hat{\varepsilon}_\perp$) as, $\mathcal{A}_{\hat{\varepsilon}} = A_0 \cos \frac{\chi}{2} \exp[i(\overline{\varphi} - \frac{\Delta}{2} - \omega t + kz)]$, $\mathcal{A}_{\hat{\varepsilon}_\perp} = A_0 \sin \frac{\chi}{2} \exp[i(\overline{\varphi} + \frac{\Delta}{2} - \omega t + kz)]$, respectively, which is spatially separated by PBS(I). The ray 1↕ travels towards ordinary mirror, M wherein it suffers ordinary reflection having reflected signal-wave, $\mathcal{A}_{\hat{\varepsilon}}^M = A_0 \cos \frac{\chi}{2} \exp[i(\overline{\varphi} - \frac{\Delta}{2} - \omega t - kz + \delta^M)]$, and the ray 2⊕ rebounds from Phase-Conjugating mirror, PCM bearing Phase-conjugated signal-wave, $\mathcal{A}_{\hat{\varepsilon}_\perp}^{PCM} = A_0 \sin \frac{\chi}{2}$

exp[i(- $\overline{\varphi}$ - $\frac{\Delta}{2}$ – ωt – kz + $\delta^{PCM}$)], where $\delta^{M}$($\delta^{PCM}$) is constant phase introduced by ordinary mirror, M ( phase-conjugating mirror, PCM). The superposition of these two reflected beams at PBS(II) generates optical-field endowed with signal-wave,

$$\mathcal{A}^{PBS(II)} = [\hat{\mathbf{\varepsilon}}\ A_0 \cos \frac{\chi}{2}\ e^{i(\overline{\varphi} - \frac{\Delta}{2} + \delta^{M})} + \hat{\mathbf{\varepsilon}}_{\perp}\ A_0 \sin \frac{\chi}{2}\ e^{i(-\overline{\varphi} - \frac{\Delta}{2} + \delta^{PCM})}]\ e^{-i(\omega t + kz)}.$$

Obviously the optical field at output of PBS(II) is the 'essentially' single-mode and in HOPS for the 'ratio of amplitudes' $\frac{A_{0\hat{\varepsilon}_{\perp}}}{A_{0\hat{\varepsilon}}}$, is $\tan\frac{\chi}{2}$, a non-random parameter, and 'sum in phases' $\varphi_{\hat{\varepsilon}_{\perp}} + \varphi_{\hat{\varepsilon}}$, is -Δ, a non-random parameter, on suitably adjusting $\delta^{M}$ and $\delta^{PCM}$. Effectively, PCMI performs conversion of linearly polarized 'truly' single-mode light into 'essentially' single-mode hidden optical-polarized light and thereby hiding ordinary optical-polarization. Quantum theory of PCMI can, simply, be given by adopting quantum-mechanical operator for positive-frequency part of vector potential, Eq.(9) and following equivalent arguments.

**4 Hidden Optical-Polarization Parameters and their Measurements**

Optical-polarization states in Classical Optics is characterized by Stokes Parameters which are quantum mechanical expectation values of their hermitian counterparts in quantum optics defined [40] by,

$$\hat{S}_o = \hat{a}^{\dagger}_y(t)\ \hat{a}_y(t) + \hat{a}^{\dagger}_x(t)\ \hat{a}_x(t)$$

$$\hat{S}_1 = \hat{a}^{\dagger}_y(t)\ \hat{a}_y(t) - \hat{a}^{\dagger}_x(t)\ \hat{a}_x(t)$$

$$\hat{S}_2 + i\hat{S}_3 = 2\ \hat{a}^{\dagger}_y(t)\ \hat{a}_x(t) \qquad (16)$$

$\hat{a}_{x,y}(t)$ gives quantum complex amplitudes of optically-polarized field-modes in the basis ($\hat{\mathbf{e}}_x$, $\hat{\mathbf{e}}_y$). Taking non-random vanishing angle parameters ($\chi_h = 0 = \Delta_h$) and the basis of description ($\hat{\mathbf{\varepsilon}}$, $\hat{\mathbf{\varepsilon}}_{\perp}$) as the linear-polarization basis ($\hat{\mathbf{e}}_x$, $\hat{\mathbf{e}}_y$) in Eq. (8) for evaluation of Classical Stokes Parameters, for HOPS, noting the fact that random variables φ has equal probability between 0 to 2π, one obtains,

$$s_0 = A_0{}^2 \text{ and } s_1 = s_2 = s_3 = 0. \tag{17}$$

Obviously, Eq.(17), at first glance, demonstrates that the light is in Unpolarized state which is not, strictly, the fact because light is in HOPS, Eq.(8). Several investigators [41-42] have critically showed inadequacies of Stokes-Parameters in characterizing optical-polarization state. An alternative characterization of perfect optical-polarization state is provided by Singh and Prakash [10].

The polarization properties of an essentially 'single-mode' electromagnetic field state such as HOPS may be described by introducing Hidden Optical –Polarization Parameters, defined in Classical optics, by,

$$h_0 = <|\underline{A}_y|^2 + |\underline{A}_x|^2>,$$

$$h_1 = <|\underline{A}_y|^2 - |\underline{A}_x|^2>,$$

$$h_2 + i h_3 = 2<\underline{A}_y \underline{A}_x>, \tag{18}$$

or, in Quantum Optics by,

$$\hat{H}_o = \hat{a}^\dagger{}_y(t)\hat{a}_y(t) + \hat{a}^\dagger{}_x(t)\hat{a}_x(t),$$

$$\hat{H}_1 = \hat{a}^\dagger{}_y(t)\hat{a}_y(t) - \hat{a}^\dagger{}_x(t)\hat{a}_x(t), \tag{19}$$

$$\hat{H}_2 + i \hat{H}_3 = 2 e^{2i\omega t} \hat{a}_y(t)\hat{a}_x(t),$$

having quantum mechanical expectation values in optical-field states, $\rho(0)$,

$$h_0 = <\hat{H}_o> = \mathrm{Tr}[\rho(0)\{\hat{a}^\dagger{}_y(t)\hat{a}_y(t) + \hat{a}^\dagger{}_x(t)\hat{a}_x(t)\}],$$

$$h_1 = <\hat{H}_1> = \mathrm{Tr}[\rho(0)\{\hat{a}^\dagger{}_y(t)\hat{a}_y(t) - \hat{a}^\dagger{}_x(t)\hat{a}_x(t)\}],$$

$$h_2 + i h_3 = <\hat{H}_2 + i \hat{H}_3> = 2 e^{2i\omega t} \mathrm{Tr}[\rho(0)\hat{a}_y(t)\hat{a}_x(t)], \tag{20}$$

where the factor $e^{2i\omega t}$ introduces only movement in phase space yet provides coherence function as in Eq.(15), and hence can be discarded for measurement purposes. The Hidden Optical-Polarization operators, Eq.(19) are seen to obey SU(2) Lie algebra,

$$[\hat{H}_1, \hat{H}_0] = [\hat{H}_1, \hat{H}_2] = [\hat{H}_1, \hat{H}_3] = 0$$

$$[\hat{H}_0, \hat{H}_2] = 2i\hat{H}_3, [\hat{H}_0, \hat{H}_3] = 2i\hat{H}_2$$

$$[\hat{H}_2, \hat{H}_3] = 2i\,(\mathbb{1}+\hat{H}_0) \qquad (21)$$

having relationship $\hat{H}_1^2 + \hat{H}_2^2 + \hat{H}_3^2 = \hat{H}_0^2 + 2\,(\mathbb{1} + \hat{H}_0)$ or $\overline{\hat{H}}^2 - \hat{H}_0^2 = 2(\mathbb{1} + \hat{H}_0)$, where $\mathbb{1}$ is identity operator. Comparing Eq.(21) with SU(2) Lie group algebraic equations of Stokes operators,

$$[\hat{S}_0, \hat{S}_1] = [\hat{S}_0, \hat{S}_2] = [\hat{S}_0, \hat{S}_3] = 0;\ [\hat{S}_1, \hat{S}_2] = 2i\hat{S}_3,\ [\hat{S}_2, \hat{S}_3] = 2i\hat{S}_1,\ [\hat{S}_3, \hat{S}_1] = 2i\hat{S}_2 \qquad (20)$$

one may take cognizance that hidden-polarization operator $\hat{H}_1$ commutes with all other parameters while $\hat{H}_0$ not contrary to $\hat{S}_0$. which is distinctive and contrasting feature of Hidden Optical-Polarization operator clearly visible in the study of squeezing in HOPS. Non-commutability of Hidden optical-polarization parameter precludes their simultaneous measurements. Heisenberg Uncertainty Principle ($\Delta\hat{H}_j^2\,\Delta\hat{H}_k^2 \geq \left|\frac{1}{2i}\langle[\hat{H}_j,\hat{H}_k]\rangle\right|^2$) can be invoked to investigate for Hidden Optical-Polarization squeezing in Degenerate Parametric Amplification [34-35].

The parameters $\hat{H}_0$, $\hat{H}_1$ are obtained by simultaneous counting of photons in two orthogonally-polarized components spatially separated by PBS. The statistical properties of the Hidden Optical-Polarization Parameters $\hat{H}_0$, $\hat{H}_1$ may be obtained from the spectrum analyzer of which provides sum and difference of detected photon numbers. The formal experimental setup for their measurement may be sketched as in fig.2.

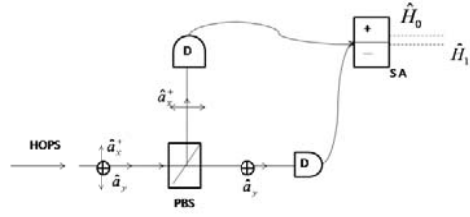

**Fig.2** Formal Proposed experiment for measurements of the Hidden Optical-Polarization Parameter $H_0$, $H_1$; PBS(Polarization Beam Splitter), D(Detector), SA(spectrum analyzer).

The measurement of $\hat{H}_2$-parameter is accomplished by allowing HOPS to pass through a rotator inclined with its optic axis at $45^0$ with respect to the x axis which converts the basis of description ($\hat{e}_x$, $\hat{e}_y$) to new primed description of bases ($\hat{e}_{x'}$, $\hat{e}_{y'}$) so that quantum complex amplitudes get transformed as,

$\hat{a}^\dagger_{x'} = (2)^{-1/2}(\hat{a}^\dagger_x + \hat{a}_y)$ and $\hat{a}_{y'} = (2)^{-1/2}(-\hat{a}^\dagger_x + \hat{a}_y)/2^{1/2}$ (see Eq.10). Rotated beam is sent into the PBS (see Fig 4) which separated spatially it into their orthogonal components in the basis ($\hat{e}_{x'}$, $\hat{e}_{y'}$). $\hat{H}_2$ - parameter of the incident beam is now obtained by taking the difference-signal by SA between detected photon numbers by the detectors, D owing to simple relationship, $\hat{H}_2 = \hat{a}_y\hat{a}_x + \hat{a}^\dagger_y\hat{a}^\dagger_x = \hat{a}^\dagger_{x'}\hat{a}_{x'} - \hat{a}^\dagger_{y'}\hat{a}_{y'}$.

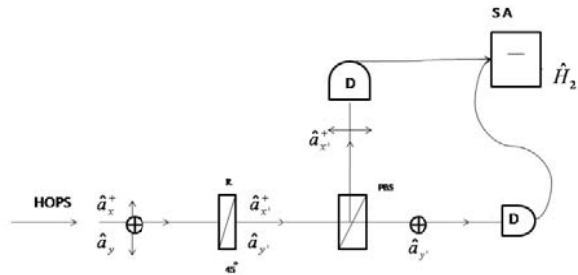

**Fig.3** Formal Proposed experiment for measurement of the Hidden Polarization Parameter $H_2$; R(rotator), PBS(Polarization Beam Splitter), D(Detector), SA(spectrum analyzer). Clearly, the difference in photon-numbers in the basis ($\hat{e}_{x'}$, $\hat{e}_{y'}$) is the parameter $\hat{H}_2$ in the basis ($\hat{e}_x$, $\hat{e}_y$).

Measurement of the $\hat{H}_3$-parameter is affected by a slight variant of the above method in which a phase-shifter is inserted before rotator on the way of HOPS which is then allowed to impinge on PBS. Effectively, the basis ($\hat{\mathbf{e}}_x$, $\hat{\mathbf{e}}_y$) is transformed to new basis ($\hat{\mathbf{e}}_{x'}$, $-i\hat{\mathbf{e}}_{y'}$) so that quantum complex amplitudes bear relations $\hat{a}_{x'}^\dagger = (2)^{-1/2}(\hat{a}_x^\dagger - i\hat{a}_y)$ and $\hat{a}_{y'} = -(2)^{-1/2}(\hat{a}_x^\dagger + i\hat{a}_y)$ (see Eq.10). Thus the $\hat{H}_3$-parameter of the beam is obtained by taking the difference-signal between detected photons by spectrum analyzer as $\hat{H}_3$ has relation, $\hat{H}_3 = i(\hat{a}_y^\dagger \hat{a}_x^\dagger - \hat{a}_y \hat{a}_x) = \hat{a}_{x'}^\dagger \hat{a}_{x'} - \hat{a}_{y'}^\dagger \hat{a}_{y'}$ (see Fig.4). Equivalent experimental proposals have been propounded by N. Korolkova et. al. for the measurements of quantum Stokes Parameters for ordinary polarized light.

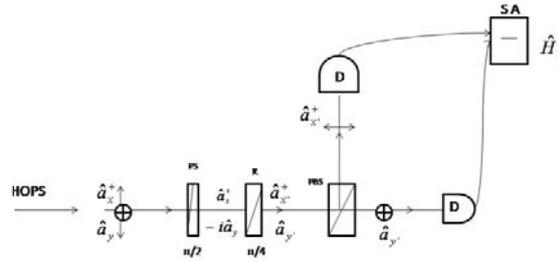

**Fig.4** Formal Proposed experiment for measurement of the Hidden Polarization Parameter $H_3$; PS(phase shifter), R(rotator), PBS(Polarization Beam Splitter), D(Detector), SA(spectrum analyzer).

**Conclusion**

We have proposed the design of a hybrid-device consituted by non-linear (Phase-conjugating) and linear-optical elements (Polarizer, Polarization Beam splitter and mirror) for the generation of Hidden optical-polarization in which 'ratio of amplitudes' and 'sum in phases' in two orthogonal modes are non-random parameters in contrast to the usual optical-polarization where 'ratio of amplitudes' and 'difference in phases' preserves non-random values. It is seen that usual optical-polarization can be concealed by inter-converting it into HOPS. HOPS is not characterized by Stokes parameters and, therefore, Hidden optical-polarization parameters are introduced and proposals for their measurements are described. HOPS, being not a polarized-state in ordinary sense, may serve generalization of Optical-Polarization state furnishing an addendum to the electromagnetic field states.

.